\documentclass[letterpaper,english, showpacs, showkeys, preprint]{revtex4}
\usepackage{amssymb}
\usepackage{graphicx}
\usepackage{epsf}
\usepackage{epsfig}

\begin{document}

\title{Relativistic Motion of Spinning Particles \\
in a Gravitational Field}

\author{C. Chicone}

\address{Department of Mathematics, University of Missouri-Columbia, Columbia,
Missouri 65211, USA}

\author{B. Mashhoon\footnote[1]{Corresponding author. E-mail
address: mashhoonb@missouri.edu (B. Mashhoon).}}

\address{Department of Physics and Astronomy, University of Missouri-Columbia,
Columbia, Missouri 65211, USA}

\author{B. Punsly}

\address{4014 Emerald Street, No. 116, Torrance, California 90503, USA;\\
International Center for Relativistic Astrophysics, University of Rome ``La
Sapienza", I-00185 Roma, Italy}

\begin{abstract}
The relative motion of a classical relativistic spinning test particle
is studied with respect to a nearby free test particle in the gravitational
field of a rotating source. 
The effects of the spin-curvature coupling force are elucidated 
and the implications of the results for the
motion of rotating plasma clumps in astrophysical jets are discussed. 
\end{abstract}


\maketitle

\section{introduction\label{sec:1}}

The purpose of this Letter is to discuss the relative motion of a spinning test particle in a gravitational field with
respect to a nearby geodesic observer. To compare the theory with observation, it proves useful to express such relative
motion in a quasi-inertial Fermi normal coordinate system that can be set up along the worldline of the reference
observer
\cite{key-1}. This approach has been recently employed for the general motion of a test particle; in particular,
astrophysically significant results have been obtained for ultrarelativistic relative motion, i.e. motion with relative
speed above a critical speed given by $c/\sqrt{{2}}\simeq0.7c$ \cite{key-2}. The present work aims to extend previous
results to the case of a spinning test mass. We choose units such that $c=1$ 
in the rest of this Letter.

The motion of an extended body in a gravitational field can be described
using the Mathisson-Papapetrou-Dixon equations \cite{key-3}. If we
neglect the quadrupole and higher moments of the particle, then the
Mathisson-Papapetrou ({}``pole-dipole'') equations suffice; these
are given by 
\begin{eqnarray}
\frac{DP^{\mu}}{d\tau} & = &-\frac{1}{2}R_{\;\;\nu\alpha\beta}^{\mu}u^{\nu}S^{\alpha\beta},\label{eq:1}\\
\frac{DS^{\mu\nu}}{d\tau} & = &P^{\mu}u^{\nu}-P^{\nu}u^{\mu}.\label{eq:2}
\end{eqnarray}
 Here $\tau$ is the proper time, $u^{\mu}=dx^{\mu}/d\tau$ is the
four-velocity, $P^{\mu}$ is the four-momentum of the particle and
$S^{\mu\nu}$ is its spin tensor. Suppose that the particle has mass
$m$ and spin $s_{0}$ and is moving in the gravitational field of
an astronomical mass $M\gg m$. In general, the dipole interaction
must be much smaller than the monopole interaction in our approximation
scheme; thus, the spin-curvature force in equation (\ref{eq:1}),
$\sim(GM/r^{3})s_{0}$, must be much smaller than the Newtonian force
$(GMm/r^{2})$. This means that the M\o ller radius~\cite{key-4}  of the 
particle, $\rho:=s_{0}/m$, should be much smaller than the distance
$r$ between $m$ and $M$.

Equations (\ref{eq:1}) and (\ref{eq:2}) must be supplemented with
additional constraints on the spin tensor. For extended bodies, the
appropriate condition turns out to be \cite{key-5}
\begin{equation}
S^{\mu\nu}P_{\nu}=0.\label{eq:3}
\end{equation}

In section \ref{sec:2}, we develop an iterative scheme for the solution
of equations (\ref{eq:1})-(\ref{eq:3}). Section \ref{sec:3} is
devoted to a discussion of the spin-curvature force. The equations
of relative motion are developed in section \ref{sec:4} and specialized
to the exterior Kerr spacetime in section \ref{sec:5}. 
The effects of spin-curvature coupling along the rotation axis of a Kerr
black hole are of particular relevance to current theoretical speculation on
the nature of gamma ray bursts.
We conclude
with a brief discussion of our results in section \ref{sec:6}.

\section{Approximation scheme\label{sec:2}}

It is necessary to develop an iterative approximation method for the
solution of equations (\ref{eq:1})-(\ref{eq:3}). We note that these
equations imply that $u_{\mu}DP^{\mu}/d\tau=0$ and 
\begin{equation}
\frac{1}{2}S_{\mu\nu}S^{\mu\nu}=s_{0}^{2},\label{eq:4}
\end{equation}
where $s_{0}$ is the constant magnitude of the spin of the particle.
Differentiating (\ref{eq:3}) with respect to proper time and using
(\ref{eq:2}) results in
\begin{equation}
(P\cdot u)P^{\mu}-P^{2}u^{\mu}+S^{\mu\nu}\frac{DP_{\nu}}{d\tau}=0.\label{eq:5}
\end{equation}
Next, we multiply (\ref{eq:5}) by $DP_{\mu}/d\tau$ and use the antisymmetry
of the spin tensor to get
\begin{equation}
(P\cdot u)\frac{DP^{2}}{d\tau}=0.\label{eq:6}
\end{equation}
Thus either $P\cdot u=0$ or $P^{2}=-m^{2}$, where $m$ has the interpretation
of the constant mass of the spinning particle. The signature of the
spacetime metric is assumed to be $+2$ throughout this work. The
multipole approximation method discussed in the previous section implies
that $P^{\mu}$ should be approximately parallel to $u^{\mu}$ except
for small corrections due to the spin of the particle; in fact, multiplication
of (\ref{eq:2}) with $u_{\nu}$ results in
\begin{equation}
P^{\mu}=-(P\cdot u)u^{\mu}-u_{\nu}\frac{DS^{\mu\nu}}{d\tau}.\label{eq:7}
\end{equation}
 We therefore assume that $P\cdot u\neq0$ and write
\begin{equation}
P^{\mu}=mu^{\mu}+E^{\mu},\label{eq:8}
\end{equation}
where the extra term $E^{\mu}$ is the small contribution of the spin
to the canonical momentum. From (\ref{eq:8}), we find $m=-P\cdot u+E\cdot u$;
on the other hand, multiplying (\ref{eq:5}) with $u_{\mu}$ results
in
\begin{equation}
m^{2}=(P\cdot u)^{2}+u_{\mu}S^{\mu\nu}\frac{DP_{\nu}}{d\tau}.\label{eq:9}
\end{equation}
It thus follows from equations~(\ref{eq:8}) and~(\ref{eq:9})
that $E\cdot u=m(1-\sqrt{1-\epsilon})$, where
\begin{equation}
\epsilon=\frac{1}{m^{2}}u_{\mu}S^{\mu\nu}\frac{DP_{\nu}}{d\tau}\label{eq:10}
\end{equation}
is $\sim(GM/r)(\rho/r)^{2}\ll 1$.

Differentiating (\ref{eq:8}) with respect to $\tau$ and using (\ref{eq:1})
we get
\begin{equation}
m\frac{Du^{\mu}}{d\tau} = -\frac{1}{2}R_{\;\;\nu\alpha\beta}^{\mu}u^{\nu}S^{\alpha\beta}-\frac{DE^{\mu}}{d\tau}.\label{eq:11}
\end{equation}
To determine $E^{\mu}$, we substitute (\ref{eq:8}) in (\ref{eq:5})
to get 
\begin{equation}
\sqrt{1-\epsilon}\,E^{\mu}=(E\cdot u)u^{\mu}+\frac{1}{m}S^{\mu\nu}\frac{DP_{\nu}}{d\tau}.\label{eq:12}
\end{equation}
It is clear from equations (\ref{eq:10})-(\ref{eq:12}) that $E^{\mu}$
is generally small and of order $m\epsilon$; moreover, 
$DE^\mu/d\tau$ is a force that  is of second order in the spin of the particle. 
In this way, it is possible to develop an iterative scheme based on the 
small parameter $\rho/r\ll 1$. 
Restricting
our treatment to effects that are of first order in the spin, we drop
$DE^{\mu}/d\tau$ in (\ref{eq:11}) and focus on the Mathisson-Papapetrou
spin-curvature force
\begin{equation}
f ^{\mu}=-\frac{1}{2}R_{\;\;\nu\alpha\beta}^{\mu}u^{\nu}S^{\alpha\beta}.\label{eq:13}
\end{equation}

It is useful to define the spin vector $S^{\mu}$ of the extended
test particle in general as
\begin{equation}
S_{\mu}=\frac{1}{2m}\eta_{\nu\mu\rho\sigma}P^{\nu}S^{\rho\sigma},\label{eq:14}
\end{equation}
 where $\eta_{\alpha\beta\gamma\delta}=\sqrt{-g}\,\epsilon_{\alpha\beta\gamma\delta}$
is the Levi-Civita tensor and $\epsilon_{\alpha\beta\gamma\delta}$
is the alternating symbol with $\epsilon_{0123}=1$. Equation (\ref{eq:14})
implies that $S_{\mu}S^{\mu}=s_{0}^{2}$; moreover, 
\begin{equation}
mS^{\alpha\beta}=\eta^{\alpha\beta\gamma\delta}P_{\gamma}S_{\delta}.\label{eq:15}
\end{equation}
It follows from the results of this section that to linear order in
spin, one can replace $P^{\mu}$ by $mu^{\mu}$ in equations (\ref{eq:14})
and (\ref{eq:15}).

\section{Spin-curvature force\label{sec:3}}

Consider the linear post-Newtonian gravitational field of a source
with mass $M$ and angular momentum $J$ given by
\begin{equation}
-ds^{2}=-(1+2\Phi)dt^{2}-4(\mathbf{A}\cdot d\mathbf{x})dt+(1-2\Phi)\delta_{ij}dx^{i}dx^{j},\label{eq:16}
\end{equation}
where $\Phi=-GM/r$ is the Newtonian gravitoelectric potential, 
$\mathbf{A}=G\mathbf{J}\times\mathbf{x}/r^{3}$
is the gravitomagnetic vector potential and $r=|\mathbf{x}|$. The
gravitomagnetic field is given by $\mathbf{B}=\mathbf{\nabla}\times \mathbf{A}$
in analogy with electrodynamics. If an ideal torque-free test gyroscope is placed
at $\mathbf{x}$, its precession frequency is given by 
\begin{equation}
\mathbf{B}=\frac{G}{r^{5}}[3(\mathbf{x}\cdot\mathbf{J})\mathbf{x}
-\mathbf{J}r^2],\label{eq:17}
\end{equation}
so that the gravitomagnetic field has the interpretation of a precession
frequency in conformity with the gravitational Larmor theorem. 
Expressing
the metric tensor in (\ref{eq:16}) as $g_{\mu\nu}=\eta_{\mu\nu}+h_{\mu\nu}$, we note
that $h_{00}=-2\Phi$, $h_{ij}=-2\Phi\delta_{ij}$ and $h_{0i}=-2A_{i}$.
The corresponding Riemann curvature tensor is given by 
\begin{equation}
R_{\mu\nu\rho\sigma}=\frac{1}{2}(h_{\mu\sigma,\nu\rho}+ h_{\nu\rho,\mu\sigma}- h_{\nu\sigma,\mu\rho}- h_{\mu\rho,\nu\sigma}).\label{eq:18}
\end{equation}

Imagine a free test particle $\mathcal{\mathcal{S}}$ initially at
rest in the stationary gravitational field of the source. The particle
carries an orthonormal tetrad frame $\tilde \lambda_{\;\;(\alpha)}^{\mu}$, 
where
$\tilde \lambda_{\;\;(0)}^{\mu}=u^{\mu}$ is its local temporal axis and
$\tilde\lambda_{\;\;(i)}^{\mu}$, $i=1,2,3$, are its local spatial axes.
The curvature tensor as measured by $\mathcal{\mathcal{S}}$ is given
by the projection of (\ref{eq:18}) on its local tetrad frame. To
calculate this curvature in the linear approximation in the gravitational 
potentials, one may set
$\tilde\lambda_{\;\;(\alpha)}^{\mu}=\delta_{\;\;\alpha}^{\mu}$; then, the
measured components of the Riemann tensor can be expressed as a $6\times6$
matrix
\begin{equation}
\left[\begin{array}{cc}
\mathcal{E} & \mathcal{B}\\
\mathcal{B} & -\mathcal{E}\end{array}\right]\label{eq:19}
\end{equation}
with indices that range over $\{01,02,03,23,31,12\} $. 
Here $\mathcal{E}$ and $\mathcal{B}$
are the electric and magnetic components of the curvature, respectively,
and are $3\times3$ symmetric and traceless matrices given by
\begin{eqnarray}
\mathcal{E}_{ij}&= & \frac{GM}{r^{3}}(\delta_{ij}-3\hat{x}^{i}\hat{x}^{i}),\label{eq:20}\\
\mathcal{B}_{ij}&= & -3\frac{GJ}{r^{4}}[\hat{x}^{i}\hat{J}^{j}
+\hat{x}^{j}\hat{J}^{i}+ (\delta_{ij}
-5 \hat{x}^{i}\hat{x}^{j}) \hat{\mathbf{x}}\cdot\hat{\mathbf{J}}],\label{eq:21}
\end{eqnarray}
where $\hat{\mathbf{x}}=\mathbf{x}/r$ and $\hat{\mathbf{J}}=\mathbf{J}/J$.

The spin-curvature force experienced by the extended
test particle $\mathcal{S}$
is given by $f ^{(\alpha)}=f ^{\mu}\tilde\lambda_{\mu}^{\;\;(\alpha)}$,
so that using equation (\ref{eq:13}),
\begin{equation}
f ^{(0)}=0,\quad f ^{(i)}=\mathcal{B}^{ij}s_{j},\label{eq:22}
\end{equation}
where $S_{(\alpha)}=(0,\mathbf{s)}$. More explicitly, this force
can be written as
\begin{equation}
\mathbf{f} 
=-\mathbf{\nabla}(\mathbf{s}\cdot\mathbf{B}),\label{eq:23}
\end{equation}
which is the gravitational analog of the Stern-Gerlach force.

As an
illustration, let us suppose that $\mathcal{S}$ is initially at rest
on the $z$ axis with its spin along the $z$ direction, which we
take to be the rotation axis of the source; then,
\begin{equation}
\mathbf{f} =\frac{6GJs_{0}}{z^{4}}\hat{\mathbf{z}},\label{eq:24}
\end{equation}
where $z\gg2GM$. 
This Mathisson-Papapetrou force is repulsive (attractive) if $\mathcal{S}$
spins in the same (opposite) sense as the rotation of the central source.
Moreover, it is interesting to note that this force is always smaller than
the Newtonian force of attraction ($G M m /z^2$). Indeed, in realistic
astrophysical situations, the Mathisson-Papapetrou dipole force would be
very small in comparison with the Newtonian monopole force. This holds in
general relativistic situations as well, as demonstrated in 
section~\ref{sec:5}.

If instead of being at rest, $\mathcal{S}$ is initially
boosted with speed $v_0$ along the $z$ axis, then the spin-curvature
force $F ^{(\alpha)}$ experienced by the boosted particle can
be obtained via a Lorentz transformation. In fact, the spin tensor
remains invariant under a boost along the $z$ axis, so that 
$S^{12}=-S^{21}=s_0$
are the only nonzero components of the spin tensor. As discussed in
the next section, the curvature tensor remains invariant as well.
In any case, the nonzero components of $F ^{(\alpha)}$ can be
simply calculated to be
\begin{equation}
F ^{(0)}=\gamma_{0}v_0f ,\quad F ^{(3)}=\gamma_{0}f ,\label{eq:25}
\end{equation}
where $\gamma_0=(1-v_0^2)^{-1/2}$ and 
$f =6GJs_{0}/z^{4}$ is given by equation~(\ref{eq:24}).

It should be emphasized that equations~(\ref{eq:22})--(\ref{eq:25}) 
are valid only to
first order in the spin of the extended test particle. Similar results have
been derived before using the other principal interpretation of the
Mathisson-Papapetrou equations, namely, within the context of a classical
point particle with ``intrinsic'' spin; that is, a point gyro that satisfies
instead of (\ref{eq:3}) the Pirani supplementary conditions 
$S^{\mu \nu} u_\nu = 0$ \cite{new6}.

\section{Relative motion in Fermi coordinates\label{sec:4}}

Imagine a free test ``observer'' $\mathcal{O}$ following
a timelike geodesic in the exterior gravitational field of an astronomical
source. Let $\lambda^{\mu}_{\;\;\;(\alpha)}$ be an orthonormal tetrad frame
that is parallel transported along the worldline of $\mathcal{O}$.
A Fermi normal coordinate system $(T,\mathbf{X})$ can be set up in
the neighborhood of this worldline based on $\lambda_{\;\;(\alpha)}^{\mu}$
as the local axes such that $\mathcal{O}$ remains at the spatial
origin of this coordinate system. Thus $\mathcal{O}$ has
Fermi coordinates $(T,\mathbf{0})$, where $T$ is the proper time of 
$\mathcal{O}$. We need to
express the motion of a spinning test particle $\mathcal{S}$
\begin{equation}
\frac{Du^{\mu}}{d\tau}=\mathcal{A}^{\mu},\quad\mathcal{A}^{\mu}\approx-\frac{1}{2m}R_{\;\;\nu\alpha\beta}^{\mu}u^{\nu}
S^{\alpha\beta}, \label{eq:26}
\end{equation}
with respect to $\mathcal{O}$. The equation of relative motion has
been derived in general in \cite{key-6} and is given by
\begin{equation}
\frac{d^{2}X^{i}}{dT^{2}}+(\Gamma_{\alpha\beta}^{i}-\Gamma_{\alpha\beta}^{0}V^{i})\frac{dX^{\alpha}}{dT}
\frac{dX^{\beta}}{dT} =\frac{1}{\Gamma^{2}}(\mathcal{A}^{i}-\mathcal{A}^{0}V^{i}),\label{eq:27}
\end{equation}
where the four-velocity $u^{\mu}$ in Fermi coordinates is $\Gamma(1,\mathbf{V})$
with $\Gamma=dT/d\tau$. The requirement that the worldline of $\mathcal{S}$
be timelike can be expressed as
\begin{equation}
\Gamma^{-2}=-g_{00}-2g_{0i}V^{i}-g_{ij}V^{i}V^{j}>0.\label{eq:28}
\end{equation}
The metric in Fermi coordinates is given by 
\begin{eqnarray}
g_{00}&= & -1-\;^{F}R_{0i0j}(T)X^{i}X^{j}+\cdots,\nonumber \\
g_{0i}&= & -\frac{2}{3}\;^{F}R_{0jik}(T)X^{j}X^{k}+\cdots,\nonumber \\
g_{ij}&= & \delta_{ij}-\frac{1}{3}\;^{F}R_{ikjl}(T)X^{k}X^{l}+\cdots,\label{eq:29}
\end{eqnarray}
where
\begin{equation}
\;^{F}R_{\alpha\beta\gamma\delta}(T)=R_{\mu\nu\rho\sigma}\lambda_{\;\;(\alpha)}^{\mu}\lambda_{\;\;(\beta)}^{\nu}
\lambda_{\;\;(\gamma)}^{\rho} \lambda_{\;\;(\delta)}^{\sigma}.\label{eq:30}
\end{equation}
The Fermi coordinates are admissible in a cylindrical region with
$|\mathbf{X}|<\mathcal{R}$ along the reference worldline such that
$\mathcal{R}(T)$ is a certain minimum radius of curvature of spacetime.

In the next section, equation (\ref{eq:27}) is studied for the case
of motion along the rotation axis of a Kerr source. The 
motion of spinning test particles in the Kerr field has been the subject
of previous investigations \cite{key-7,key-8,key-9,key-10,key-11,key-12};
however, in the present work we consider the deviation of such motion
relative to a reference geodesic in the context of a Fermi normal
coordinate system that is constructed along the reference geodesic.
This approach turns out to be of direct observational relevance for
the motion of clumps in astrophysical jets as discussed in section~\ref{sec:6}.

\section{Motion in the Kerr field\label{sec:5}}

The purpose of this section is to study the motion of $\mathcal{S}$
relative to $\mathcal{O}$ along the rotation axis of a Kerr source.
The Kerr metric is given by
\begin{eqnarray}
-ds^{2}&= & -dt^{2}+\Sigma\left(\frac{1}{\Delta}dr^{2}+d\theta^{2}\right)+(r^{2}+a^{2})\sin^{2}\theta d\phi^{2}\nonumber \\
 && \quad+2GM\frac{r}{\Sigma}(dt-a\sin^{2}\theta d\phi)^{2}\label{eq:31}
\end{eqnarray}
in Boyer-Lindquist coordinates, where $\Sigma=r^{2}+a^{2}\cos^{2}\theta$
and $\Delta=r^{2}-2GMr+a^{2}$. Here $M$ and $a>0$ are respectively
the mass and the specific angular momentum ($J/M$) of the source. We assume
that the free reference particle $\mathcal{O}$ moves along the rotation
axis on an escape trajectory. The geodesic equations of motion of
$\mathcal{O}$ reduce to
\begin{equation}
\frac{dt}{ds}=\gamma\frac{r^{2}+a^{2}}{r^{2}-2GMr+a^{2}},\quad \frac{dr}{ds}
=\sqrt{\gamma^{2}-1+\frac{2GMr}{r^{2}+a^{2}}}.\label{eq:32}
\end{equation}
Here $\gamma\geq1$ is a constant of integration such that for $r\to\infty,\gamma$
is the Lorentz factor of the particle as measured by the static inertial
observers at spatial infinity. We integrate system (\ref{eq:32})
with the initial conditions that at $s=0$, $t=0$ and $r=r_{0}>\sqrt{3}a$.
Moreover, the spinning test particle $\mathcal{S}$ also starts from
this same event with speed $V_{0}>0$ relative to $\mathcal{O}$ and
moves along the rotation axis. To describe this relative motion, we
establish a Fermi coordinate system along the worldline of $\mathcal{O}$.
In $(t,r,\theta,\phi)$ coordinates, the orthonormal tetrad frame
$\lambda^{\mu}_{\;\;(\alpha)}$ is such that
\begin{equation}
\lambda_{\;\;(0)}^{\mu}=(\dot{t},\dot{r},0,0),\quad\lambda_{\;\;(3)}^{\mu}=(\gamma^{-1}\dot{t}\dot{r},\gamma,0,0),\label{eq:33}
\end{equation}
where $\dot{t}=dt/ds$ and $\dot{r}=dr/ds$ are given by~(\ref{eq:32}).
The axial symmetry about the rotation axis implies that there is a
simple rotational degeneracy in the choice of $\lambda_{\;\;(1)}^{\mu}$
and $\lambda_{\;\;(2)}^{\mu}$. The projection of the curvature tensor
on the tetrad frame of $\mathcal{O}$, given by equation (\ref{eq:30}),
turns out to be independent of the explicit choice for $\lambda_{\;\;(1)}^{\mu}$
and $\lambda_{\;\;(2)}^{\mu}$ and may be expressed as before in terms
of $\mathcal{E}$ and $\mathcal{B}$ such that 
$\mathcal{E}/k=\mbox{diag}\left(-\frac{1}{2},-\frac{1}{2},1\right)$
and $\mathcal{B}/q=\mbox{diag}\left(-\frac{1}{2},-\frac{1}{2},1\right)$, where
\begin{equation}
k=-2\,GM\frac{r(r^{2}-3a^{2})}{(r^{2}+a^{2})^{3}},\quad q=2\,GMa\frac{3r^{2}-a^{2}}{(r^{2}+a^{2})^{3}}.\label{eq:34}
\end{equation}
Thus for a given $r$, the curvature measured by $\mathcal{O}$ is
completely independent of $\gamma$. This remarkable fact is a consequence of the
degenerate nature of the Kerr solution, which implies that the rotation
axis of the exterior Kerr spacetime corresponds to two special tidal
directions for ingoing and outgoing trajectories; in fact, $\mathcal{E}$
and $\mathcal{B}$ are invariant under a boost along the rotation
axis \cite{key-13,key-14}.
\begin{figure}
\end{figure}

The equation of motion of $\mathcal{S}$ relative to $\mathcal{O}$ is
given by equations (\ref{eq:26}) and (\ref{eq:27}), where we need
to evaluate the force term involving $R_{\;\;\nu\alpha\beta}^{\mu}(T,\mathbf{X})$
and $S^{\alpha\beta}(T,\mathbf{X})$ in the Fermi system. To simplify
matters, we assume that 
$R_{\;\;\nu\alpha\beta}^{\mu}(T,\mathbf{X})
\approx R_{\;\;\nu\alpha\beta}^{\mu}(T,\mathbf{0})$,
which can be determined by equation (\ref{eq:34}), and 
$S^{\alpha\beta}(T,\mathbf{X})\approx S^{\alpha\beta}(T,\mathbf{0})$.
As in the previous
section, the spin  of the test mass $\mathcal{S}$ is assumed
to be along the axis of rotation, so that the only nonzero components
of the spin tensor are $S^{12}=-S^{21}=s_{0}$. 
It follows that $\mathcal{A}^{0}=q\rho\Gamma V$
and $\mathcal{A}^{3}=q\rho\Gamma$ are the only nonzero components
of $\mathcal{A}^{\mu}$, where $\Gamma^{-1}=\sqrt{1-\dot{Z}^{2}+kZ^{2}}$
and $\rho= s_{0}/m$ is the M\o  ller radius of $\mathcal{S}$. With
these simplifications and taking only the terms given in (\ref{eq:29})
into account, equation (\ref{eq:27}) reduces to
\begin{equation}
\ddot{Z}+k(1-2\dot{Z}^{2})Z=q\rho(1-\dot{Z}^{2})\Gamma^{-1}.\label{eq:35}
\end{equation}
This equation takes a dimensionless form if all lengths are expressed
in units of $GM$ and $\rho$ is replaced by  $\hat{\rho}=\rho/(GM)$. 
Then, $(GM/r)\hat{\rho}=s_{0}/(mr)\ll1$
by assumption.
\begin{figure}
\centerline{\psfig{file=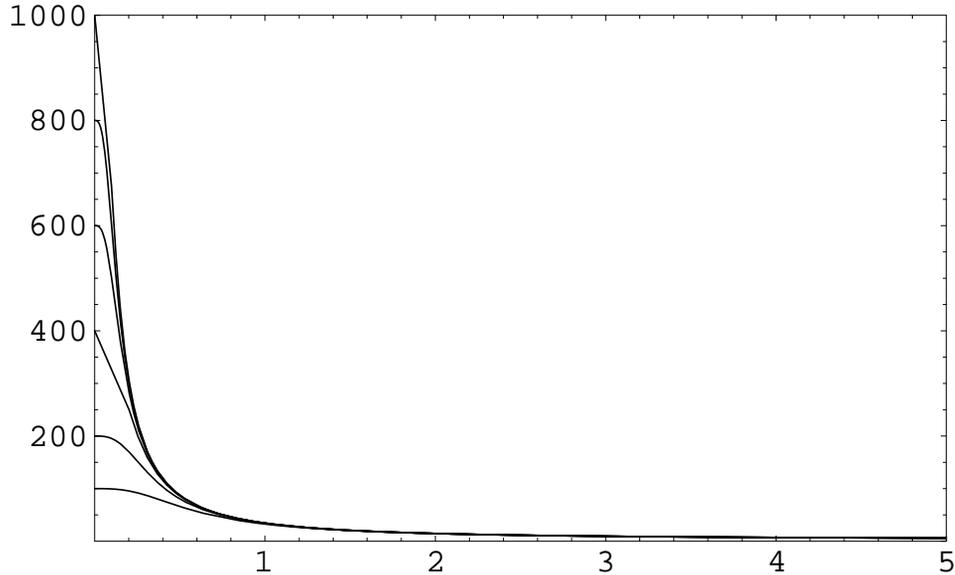, width=30pc}}
\caption{Plot of $\hat\Gamma$
versus $T/(GM)$
based on the integration of equation~(\ref{eq:35})
for $r_0=5\,GM$, $\widehat{\rho} =2$, $a=GM$ and $\Gamma_0=100, 200, 400, 600,
800$ and 1000. 
\label{fig:1}}
\end{figure}

 The term $ 1 - 2 \dot Z^2 $ in equation (\ref{eq:35}) 
is due to the fact that we
express the equation of motion of $\mathcal{S}$ with respect 
to the Fermi time $T$
rather than $\tau$, the proper time of $\mathcal{S}$. 
For $|\dot Z| \ll 1$, equation (\ref{eq:35})
reduces to a Jacobi-type equation for the relative motion of $\mathcal{S}$ with
respect to $\mathcal{O}$ \cite{key-2}.
\begin{figure}
\centerline{\psfig{file=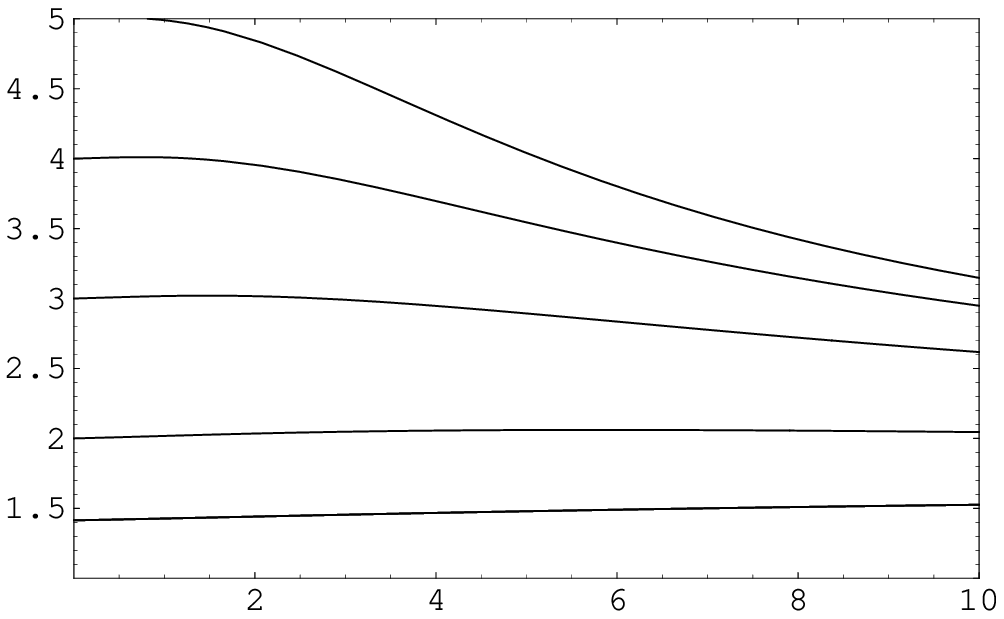, width=20pc}}
\centerline{\psfig{file=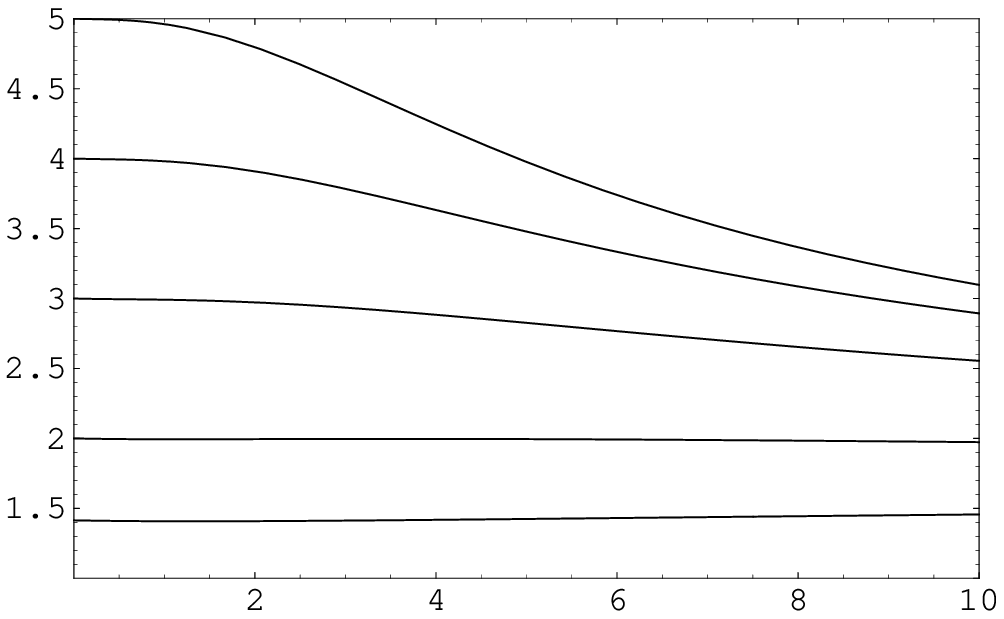,width=20pc}}
\caption{Plot of $\hat\Gamma$
versus $T/(GM)$
based on the integration of equation~(\ref{eq:35})
for $r_0=5\,GM$, $\widehat{\rho} =2$ in the top panel and $\widehat{\rho} =-2$
in the bottom panel, $a=GM$ and $\Gamma_0=\sqrt{2}, 2, 3, 4$ and 5. 
\label{fig:2}}
\end{figure}

To characterize the relative motion invariantly, let us imagine a
set of static observers in the Fermi system situated along the $Z$
axis and let $\hat{\Gamma}$ be the Lorentz factor of $\mathcal{S}$
as measured by these static Fermi observers that are in general accelerated.
Then $\hat{\Gamma}=\sqrt{-g_{00}}\;\Gamma$, where $-g_{00}=1+kZ^{2}$
in our approximation scheme; therefore, $\hat\Gamma\le \Gamma$, since
$k<0$.

Equation (\ref{eq:35}) contains the {}``electric'' and {}``magnetic''
curvatures $k$ and $q$, respectively, that can be expressed as functions
of the Fermi coordinate time $T$. That is, the equation for $\dot{r}$ in
(\ref{eq:32}) can be integrated such that with $s\mapsto T$, we
find $r(T)$ and hence $k(T)$ and $q(T)$ that must then be substituted
in (\ref{eq:35}). This system has been numerically integrated 
with initial conditions that at $T = 0$, $Z = 0$ and $\dot Z = V_0 > 0$, and
the results are presented in figures 1 and 2, where $\hat{\Gamma}$ is plotted
versus $T/(GM)$ for ultrarelativistic relative motion along the rotation
axis of a maximal $(a=GM)$ Kerr source. 
We note that with our initial conditions both $\Gamma$ and 
$\hat \Gamma$ at $T =0$ are given by $\Gamma_0 = ( 1 - V_0^2 ) ^{-1/2}$, which is the Lorentz factor
corresponding to $V_0 < 1$.
For motion along the $Z$ axis, $\hat\Gamma\le \Gamma$;
however,
it turns out that a similar
plot for $\Gamma$ would be indistinguishable from figure~\ref{fig:1}. 
We choose
$\gamma=1$ in figures 1 and 2, since the results turn out to be independent
of the choice of $\gamma$ so long as $\mathcal{O}$ is slowly outgoing.
For highly ultrarelativistic motion as in figure~\ref{fig:1}, the spin of the clump
has a negligible influence on its deceleration. In fact, with the same
conditions as in figure~\ref{fig:1} but for $\hat\rho = 0$ and 10, 
the resulting figures turn
out to be indistinguishable from figure~\ref{fig:1}.
Let us note that $\Gamma_0 \to \infty$ is not allowed here, since the back
reaction of such a particle on the gravitational field of the source cannot
be neglected and hence our test particle approximation scheme would break down.
The case where the spin of $\mathcal{S}$ is antiparallel to the $Z$
axis can be treated by formally letting $s_{0}\mapsto-s_{0}$. The
corresponding results are presented in the bottom panel of 
figure~\ref{fig:2}. The graphs in the top panel
are slightly above the corresponding ones in the bottom panel, 
though the trends are
essentially the same.

Let us suppose that the clump is homogeneous and cylindrical with radius $R$,
and the axis of the cylinder coincides with the $Z$ axis. Thus 
$s_0 = \frac{1}{2} m R^2 \omega$, where $\omega$ is the frequency of rotation 
of the clump such that 
$R\omega < 1$. Moreover, $( GM/r) \hat \rho = s_0/ (mr) \ll 1$
 implies that $\hat\rho \ll r/(GM)$.
In figures~\ref{fig:1} and~\ref{fig:2}  we have 
$ r\ge r_0 = 5\, GM $; therefore, we have
chosen $\hat \rho = 2$ to represent in effect the maximum spin of the clump.
This choice is consistent with various scenarios that have been proposed for
jet creation near the poles of a rapidly rotating Kerr black hole 
(see~\cite{key-17a}, chapters 9 and 11, and the references cited therein).

\section{Discussion\label{sec:6}}

If the initial Lorentz factor $\Gamma_{0}$ of $\mathcal{S}$ is close
to the critical value $\Gamma_{c}=\sqrt{2}$, then the spin-curvature
force is rather small but may not be negligible very close to the central black hole. 
However, for $\Gamma_{0}\gg\sqrt{2}$, the
contribution of the spin-curvature force turns out to be essentially
negligible and the spinning particle $\mathcal{S}$ decelerates toward
the critical speed.
This confirms our previous work~\cite{key-2} on the speed of jets in 
microquasars~\cite{key-15}.

In this paper,  we have shown that the current speculation~\cite{key-17} 
that the
spin-curvature coupling could be a prominent driver of astrophysical jets
from black holes is overly optimistic. We show quite generally that 
the standard tidal effects~\cite{key-2} 
will dominate the spin-curvature effects for
plasma clumps in astrophysical jets.

\end{document}